\documentstyle[twoside]{article}

\catcode`\@=11
\long\def\@makefntext#1{
\protect\noindent \hbox to 3.2pt {\hskip-.9pt  

$^{{\eightrm\@thefnmark}}$\hfil}#1\hfill}		

\def\thefootnote{\fnsymbol{footnote}}
\def\@makefnmark{\hbox to 0pt{$^{\@thefnmark}$\hss}}	

\def\ps@myheadings{\let\@mkboth\@gobbletwo
\def\@oddhead{\hbox{}
\rightmark\hfil\eightrm\thepage}   

\def\@oddfoot{}\def\@evenhead{\eightrm\thepage\hfil
\leftmark\hbox{}}\def\@evenfoot{}
\def\sectionmark##1{}\def\subsectionmark##1{}}

\oddsidemargin=\evensidemargin
\renewcommand{\thefootnote}{\fnsymbol{footnote}}

\newcounter{sectionc}\newcounter{subsectionc}\newcounter{subsubsectionc}
\renewcommand{\section}[1] {\vspace{12pt}\addtocounter{sectionc}{1} 

\setcounter{subsectionc}{0}\setcounter{subsubsectionc}{0}\noindent 

	{\tenbf\thesectionc. #1}\par\vspace{5pt}}
\renewcommand{\subsection}[1] {\vspace{12pt}\addtocounter{subsectionc}{1} 

	\setcounter{subsubsectionc}{0}\noindent 

	{\bf\thesectionc.\thesubsectionc. {\kern1pt \bfit #1}}\par\vspace{5pt}}
\renewcommand{\subsubsection}[1] {\vspace{12pt}\addtocounter{subsubsectionc}{1}
	\noindent{\tenrm\thesectionc.\thesubsectionc.\thesubsubsectionc.
	{\kern1pt \tenit #1}}\par\vspace{5pt}}
\newcommand{\nonumsection}[1] {\vspace{12pt}\noindent{\tenbf #1}
	\par\vspace{5pt}}

\newcounter{appendixc}
\newcounter{subappendixc}[appendixc]
\newcounter{subsubappendixc}[subappendixc]
\renewcommand{\thesubappendixc}{\Alph{appendixc}.\arabic{subappendixc}}
\renewcommand{\thesubsubappendixc}
	{\Alph{appendixc}.\arabic{subappendixc}.\arabic{subsubappendixc}}

\renewcommand{\appendix}[1] {\vspace{12pt}
        \refstepcounter{appendixc}
        \setcounter{figure}{0}
        \setcounter{table}{0}
        \setcounter{lemma}{0}
        \setcounter{theorem}{0}
        \setcounter{corollary}{0}
        \setcounter{definition}{0}
        \setcounter{equation}{0}
        \renewcommand{\thefigure}{\Alph{appendixc}.\arabic{figure}}
        \renewcommand{\thetable}{\Alph{appendixc}.\arabic{table}}
        \renewcommand{\theappendixc}{\Alph{appendixc}}
        \renewcommand{\thelemma}{\Alph{appendixc}.\arabic{lemma}}
        \renewcommand{\thetheorem}{\Alph{appendixc}.\arabic{theorem}}
        \renewcommand{\thedefinition}{\Alph{appendixc}.\arabic{definition}}
        \renewcommand{\thecorollary}{\Alph{appendixc}.\arabic{corollary}}
        \renewcommand{\theequation}{\Alph{appendixc}.\arabic{equation}}
        \noindent{\tenbf Appendix \theappendixc #1}\par\vspace{5pt}}
\newcommand{\subappendix}[1] {\vspace{12pt}
        \refstepcounter{subappendixc}
        \noindent{\bf Appendix \thesubappendixc. {\kern1pt \bfit #1}}
	\par\vspace{5pt}}
\newcommand{\subsubappendix}[1] {\vspace{12pt}
        \refstepcounter{subsubappendixc}
        \noindent{\rm Appendix \thesubsubappendixc. {\kern1pt \tenit #1}}
	\par\vspace{5pt}}

\newcommand{\textlineskip}{\baselineskip=13pt}
\newcommand{\smalllineskip}{\baselineskip=10pt}

\def\eightcirc{
\begin{picture}(0,0)
\put(4.4,1.8){\circle{6.5}}
\end{picture}}
\def\eightcopyright{\eightcirc\kern2.7pt\hbox{\eightrm c}}

\def\abstracts#1#2#3{{
	\centering{\begin{minipage}{4.5in}\baselineskip=10pt\footnotesize
	\parindent=0pt #1\par 

	\parindent=15pt #2\par
	\parindent=15pt #3
	\end{minipage}}\par}}

\renewenvironment{thebibliography}[1]
	{\frenchspacing
	 \ninerm\baselineskip=11pt
	 \begin{list}{\arabic{enumi}.}
	{\usecounter{enumi}\setlength{\parsep}{0pt}
	 \setlength{\leftmargin 12.7pt}{\rightmargin 0pt} 
	 \setlength{\itemsep}{0pt} \settowidth
	{\labelwidth}{#1.}\sloppy}}{\end{list}}

\newcounter{itemlistc}
\newcounter{romanlistc}
\newcounter{alphlistc}
\newcounter{arabiclistc}

\newcommand{\fcaption}[1]{
        \refstepcounter{figure}
        \setbox\@tempboxa = \hbox{\footnotesize Fig.~\thefigure. #1}
        \ifdim \wd\@tempboxa > 5in
           {\begin{center}
        \parbox{5in}{\footnotesize\smalllineskip Fig.~\thefigure. #1}
            \end{center}}
        \else
             {\begin{center}
             {\footnotesize Fig.~\thefigure. #1}
              \end{center}}
        \fi}

\newcommand{\tcaption}[1]{
        \refstepcounter{table}
        \setbox\@tempboxa = \hbox{\footnotesize Table~\thetable. #1}
        \ifdim \wd\@tempboxa > 5in
           {\begin{center}
        \parbox{5in}{\footnotesize\smalllineskip Table~\thetable. #1}
            \end{center}}
        \else
             {\begin{center}
             {\footnotesize Table~\thetable. #1}
              \end{center}}
        \fi}

\def\@citex[#1]#2{\if@filesw\immediate\write\@auxout
	{\string\citation{#2}}\fi
\def\@citea{}\@cite{\@for\@citeb:=#2\do
	{\@citea\def\@citea{,}\@ifundefined
	{b@\@citeb}{{\bf ?}\@warning
	{Citation `\@citeb' on page \thepage \space undefined}}
	{\csname b@\@citeb\endcsname}}}{#1}}

\newif\if@cghi
\def\cite{\@cghitrue\@ifnextchar [{\@tempswatrue
	\@citex}{\@tempswafalse\@citex[]}}
\def\citelow{\@cghifalse\@ifnextchar [{\@tempswatrue
	\@citex}{\@tempswafalse\@citex[]}}
\def\@cite#1#2{{$\null^{#1}$\if@tempswa\typeout
	{IJCGA warning: optional citation argument 

	ignored: `#2'} \fi}}

\def\pmb#1{\setbox0=\hbox{#1}
	\kern-.025em\copy0\kern-\wd0
	\kern.05em\copy0\kern-\wd0
	\kern-.025em\raise.0433em\box0}

\def\fnt#1#2{\footnotetext{\kern-.3em
	{$^{\mbox{\scriptsize #1}}$}{#2}}}

\def\fpage#1{\begingroup
\voffset=.3in
\thispagestyle{empty}\begin{table}[b]\centerline{\footnotesize #1}
	\end{table}\endgroup}

\def\runninghead#1#2{\pagestyle{myheadings}
\markboth{{\protect\footnotesize\it{\quad #1}}\hfill}
{\hfill{\protect\footnotesize\it{#2\quad}}}}
\font\tenrm=cmr10
\font\tenit=cmti10 

\font\tenbf=cmbx10
\font\bfit=cmbxti10 at 10pt
\font\ninerm=cmr9

\font\eightrm=cmr8

\def\qed{\hbox{${\vcenter{\vbox{			
   \hrule height 0.4pt\hbox{\vrule width 0.4pt height 6pt
   \kern5pt\vrule width 0.4pt}\hrule height 0.4pt}}}$}}

\renewcommand{\thefootnote}{\fnsymbol{footnote}}	

\def\soneg{{\cal S}_1{\cal G}}
\def\stwog{{\cal S}_2{\cal G}}

\def\bW{{\bf W}}
\def\bx{{\bf x}}
\def\by{{\bf y}}
\def\bz{{\bf z}}

\def\bp{{\bf p}}
\def\D{{\cal D}}

\def\N{{\cal N}}

\def\Tr{{\rm Tr}}
\newcommand{\ket}[1]{|#1\rangle}

\begin{document}

\runninghead{Galilean Supersymmetry in 2+1 Dimensions}  
{Oren Bergman}

\normalsize\textlineskip
\thispagestyle{empty}
\setcounter{page}{1}



\fpage{1}
\centerline{\bf GALILEAN SUPERSYMMETRY IN 2+1 DIMENSIONS\footnote{
Based on a talk given at the International Workshop on Low Dimensional
Field Theory, Telluride, Colorado, August 5--17, 1996.}}
\vspace*{0.37truein}
\centerline{\footnotesize OREN BERGMAN\footnote{Address as of September 1: 
Department of Physics, Brandeis University, Waltham, MA 02254. E-mail: 
obergman@binah.cc.brandeis.edu or bergman@string.harvard.edu.}}
\vspace*{0.015truein}
\centerline{\footnotesize\it Institute for Fundamental Theory}
\baselineskip=10pt
\centerline{\footnotesize\it Department of Physics}
\baselineskip=10pt
\centerline{\footnotesize\it University of Florida, Gainesville,
FL 32611, USA}
\vspace*{0.225truein}

\vspace*{0.21truein}
\abstracts{We review work done in collaboration with C.B. Thorn\cite{bergmantgal}
on Super-Galilei invariant field theory and its application to the
reformulation of superstring theory in terms of constituent 
superstring-bits.}{}{}


\vspace*{1pt}\textlineskip	
\section{Introduction}	
\noindent
The Galilei group is the spacetime symmetry of non-relativistic systems. It is
the low velocity limit of the Poincar\'e group, which is the
true spacetime symmetry of relativistic systems.
We are all very familiar with the success of relativistic field theories
in describing the interactions of fundamental particles.
What is the point then of studying Galilei invariant field theories?

First of all, some Galilean field theories can be understood as
non-relativistic limits of relativistic field theories. As such,
they offer a simpler setting for studying
generic field-theoretic concepts, such as
renormalization and renormalization group,\cite{bergmananomaly}
anomalies,\cite{bergmananomaly,bergmanlozano} and
solitons.\cite{jackiwpi}

Second, Galilean field theories are a second-quantized description
of quantum mechanics, and are therefore useful for addressing 
many body problems in non-relativistic (condensed matter) systems. 
A second-quantized description
often gives insight into an otherwise intractable
problem. An example of this is the Aharonov-Bohm scattering problem.
The exact solution is well known\cite{ab} and well behaved, but perturbation
theory runs into problems.\cite{pertab} Second quantization shows that 
the source of the discrepancy is a conformal anomaly arising from
the perturbative divergence of the $1/r^2$ problem, which must
be canceled by the inclusion of a contact interaction of critical
strength.\cite{bergmanlozano}

Finally, the most surprising application of Galilei invariant field 
  theories is to {\em relativistic} strings. The spacetime
  symmetry of the $D$-dimensional relativistic string in {\em light-cone} 
  gauge is the transverse $D-2$-dimensional Galilei group. 
  Only when $D=26$ does
  one regain $D$-dimensional Poincar\'e invariance. The light-cone
  string can be understood as a composite object, whose constituent
  "bits" bind to form a closed chain.\cite{thornstring} The dynamics of 
these string-bits
  are governed by a Galilei invariant matrix field theory -- a "string-bit
  model".\cite{thornbits}

Supersymmetric extensions of the Galilei group were first suggested by
Puzalowski
in $3+1$ dimensions.\cite{puzalowski} The minimal superalgebra is
called $\soneg$, and introduces a single supercharge $Q$, with $Q^2 = M$.
There is also an extended superalgebra, called
$\stwog$, that introduces another supercharge $R$, with $R^2 = H$.
 Puzalowski constructed field theoretic
representations of the minimal superalgebra $\soneg$,
which can be understood as the non-relativistic limit of an $\N=1$ 
Super-Poicar\'e algebra.
We are more interested in the extended superalgebra $\stwog$,
which 
can be understood either as the non-relativistic limit of an $\N=2$
Super-Poincar\'e algebra, or as the light-cone subalgebra of 
a one higher dimension $\N=1$ Super-Poincar\'e algebra. The first
point of view is relevant to low-energy physics,
and the second point of view is relevant to 
superstring-bits.\cite{bergmantbits}

Section 2 reviews the two-dimensional Super-Galilei algebras. Section 3
reviews the construction of Super-Galilei invariant field theories, and in section
4 we derive Non-relativistic Super-Chern-Simons theory as a special case. In 
section 5 we construct Super-Galilei matrix field theories, and discuss their
implication on a composite formulation of superstring theory. Section
6 is devoted to a discussion of the results and future directions.

\textheight=7.8truein
\setcounter{footnote}{0}
\renewcommand{\thefootnote}{\alph{footnote}}

\section{The $d=2$ Super-Galilei Algebra}
\renewcommand{\theequation}{\arabic{sectionc}.\arabic{equation}}
\noindent
To start, let us review the Galilei algebra in $d=2$ space dimensions
\footnote{Lower case $d$ will denote space dimensions only, and 
upper case $D$ will denote spacetime dimensions.}.
The generators consist of the momentum vector $P_i$, the boost vector $K_i$,
the angular momentum pseudo-scalar $J$, the Hamiltonian $H$ and the 
number operator $N$. The non-trivial part of the algebra is given by:
\begin{equation}
  [P_i,K_j] = i\delta_{ij}mN \; ,  \; [P_i,J] = -i\epsilon_{ij}P_j \; , \;
  [H,K_i] = iP_i \; ,  \; [K_i,J] = -i\epsilon_{ij}K_j \; .
\label{galilei}
\end{equation}
One usually thinks of this algebra as the non-relativistic (N-R) limit 
($c\rightarrow\infty$) of the $D=3$ Poincar\'e algebra:
\begin{eqnarray}
 ~[P^\rho,M^{\mu\nu}] &=& 
     i\big(\eta^{\nu\rho}P^\mu - \eta^{\mu\rho}P^\nu\big) \nonumber \\
 ~[M^{\mu\nu},M^{\rho\sigma}] &=& i\big(\eta^{\mu\rho}M^{\nu\sigma}
          + \eta^{\mu\sigma}M^{\rho\nu} - \eta^{\nu\rho}M^{\mu\sigma}
          - \eta^{\nu\sigma}M^{\rho\mu}\big) \; .
\end{eqnarray}
In the N-R limit the Poincar\'e generators become:
\begin{equation}
   P^0\rightarrow mN + H \qquad
   M^{0i}\rightarrow K^i \qquad
   M^{12}\rightarrow J \; ,
\end{equation}
which in turn satisfy the Galilei algebra (\ref{galilei}).
Alternatively, eq.~(\ref{galilei}) can be interpreted
as the light-cone subalgebra of the $D=4$ Poincar\'e algebra, in which 
\begin{equation}
   mN = P^0 + P^3 \quad , \quad H = (P^0 - P^3)/2 \quad , \quad
   K^i = M^{0i} + M^{3i} \quad , \quad J = M^{12} \; .
\end{equation}

Minimal supersymmetric extension of the Galilei group is achieved 
by introducing 
a complex supercharge $Q$ satisfying:
\begin{equation}
  [Q,J] = Q/2 \quad , \quad
  \{Q,Q^\dagger\} = \vphantom{1\over 2}mN \; .
 \label{Qcharge}
\end{equation}
The resulting superalgebra is called $\soneg$. 
It can be understood as the non-relativistic
limit of the $D=3$, $\N=1$ Super-Poincar\'e algebra:
\begin{equation}
  [Q_a,M^{\mu\nu}] = {1\over 2} \gamma^{\mu\nu}_{ab}Q_b \quad , \quad
 \{Q_a,Q_b\} = \left(\gamma^\mu\gamma^0\right)_{ab}P_\mu \; ,
\end{equation}
where $Q_a$ is a real two-component spinor, and
$\gamma^{\mu\nu} = i[\gamma^\mu,\gamma^\nu]/2$. The non-relativistic
limit gives $(Q_1 + iQ_2)/\sqrt{2} \rightarrow Q$.

We extend the superalgebra to $\stwog$ by introducing another supercharge
$R$, satisfying:
\begin{equation}
  [R,J] = -R/2  \; , \; [R,K^-] = iQ \; , \;
  \{R,Q^\dagger\} = -P^+/2    \; , \;  \{R,R^\dagger\} = H/2\; ,
\end{equation}
with all other (anti-)commutators vanishing.\footnote{We use the notation
$V^{\pm} = V_1 \pm iV_2$.} 
 The extended superalgebra can arise as the non-relativistic limit of the 
$D=3$, $\N=2$ Super-Poincar\'e algebra, with a non-trivial central charge.
 The relevant part of the Super-Poincar\'e algebra takes the form:
 \begin{equation}
  [Q_a,M^{\mu\nu}] = {1\over 2} \gamma^{\mu\nu}_{ab}Q_b \quad , \quad
  \{Q_a,{\overline Q}_b\} = \gamma^\mu_{ab}P_\mu - \delta_{ab}T \; ,
 \end{equation}
 where $Q_a$ is a complex two-component spinor.
 In the basis in which $\gamma^0 = \sigma_3$ the non-relativistic
 limit gives $Q_1/\sqrt{2}\rightarrow Q$ and 
$Q_2/\sqrt{2}\rightarrow R$, provided
 that the N-R limit of the central charge is $T\rightarrow -mN$.
Alternatively, $\stwog$ can be interpreted as the light-cone subalgebra of  the 
$D=4$, $\N=1$ Super-Poincar\'e algebra.
 The relevant part of the superalgebra is given by
\begin{equation}
  [Q_a,M^{\mu\nu}] = {1\over 2} \sigma^{\mu\nu}_{ab}Q_b \quad , \quad
  \{Q_a,Q^\dagger_b\} = \sigma^\mu_{ab}P_\mu \; ,
\end{equation}
where $Q_a$ is a complex two-component Weyl spinor, 
and 
$\sigma^{\mu\nu} = i(\sigma^\mu\overline{\sigma}^\nu 
- \sigma^\nu\overline{\sigma}^\mu)/2$. The Galilean supercharges are then
given by $Q = \pm Q_1 $ and $R = \mp Q_2/\sqrt{2}$.

\section{$\stwog$-Invariant Field Theory}
\renewcommand{\theequation}{\arabic{sectionc}.\arabic{equation}}
\setcounter{equation}{0}
\noindent
Introduce a complex field $\phi(\bx)$, and a complex one-component
Grassmann field $\psi(\bx)$, satisfying canonical commutation and
anti-commutation relations respectively:
\begin{equation}
 [\phi(\bx),\phi^\dagger(\by)] = \{\psi(\bx),\psi^\dagger(\by)\} =  
       \delta(\bx-\by) \; .
\end{equation}
The free field representation of the supercharges is given by:
\begin{equation}
  Q = -i\sqrt{m}\int d\bx \psi^\dagger(\bx)\phi(\bx) \qquad , \qquad
 R_0 = {1\over 2\sqrt{m}}
    \int d\bx \psi^\dagger(\bx)\partial^+\phi(\bx) \; ,
\label{freefield}
\end{equation} 
resulting in the free Hamiltonian
\begin{equation}
  H_0 = 2\{R_0,R_0^\dagger\} =
  {1\over 2m}\int d\bx \left[|\nabla\phi(\bx)|^2 + 
  |\nabla\psi(\bx)|^2\right] \; .
\end{equation}
Since the Hamiltonian of any $\stwog$-invariant field theory is the square of 
a supercharge $R$, we can construct
interacting $\stwog$-invariant field theories by adding an interaction term
to $R_0$, so that 
$R = R_0 + R^\prime$.
If we restrict to quartic
operators, the $\stwog$ algebra implies that $R^\prime$ must be of the form
\begin{equation}
 R^\prime = \int 
 d\bx\,d\by\,W^+(\by-\bx)\psi^\dagger(\bx)\rho(\by)\phi(\bx)\; ,
\label{Rprime}
\end{equation}
where $\rho = \phi^\dagger \phi + \psi^\dagger\psi$, and the {\em superpotential}
$W^+$ is given by
\begin{equation}
 W^+(\bx) = (\partial^1 + i\partial^2){\cal F}(|\bx|)
          =  \partial^+{\cal F}(|\bx|) \; ,
\end{equation}
where ${\cal F}$ is complex in general.
The total supercharge can now be written
concisely as
\begin{equation}
 R = {1\over 2{\sqrt m}}\int d\bx \psi^\dagger(\bx)\D^+\phi(\bx) \; ,
\end{equation}
where
\begin{equation}
 \D^+ = \partial^+ - i\int d\by W^+(\by - \bx)\rho(\by) \; .
\end{equation}
The Hamiltonian is obtained by squaring $R$, and is given by
\begin{eqnarray}
\lefteqn{H = {1\over 2m}\int d\bx\,\left[|\D^+\phi(\bx)|^2 + 
       |\D^+\psi(\bx)|^2\right]} \nonumber \\
  & & \mbox{} +  {1\over m}\int d\bx\,d\by\,\mbox{\boldmath $\nabla$}_y\times \bW
      (\by-\bx)\left[\psi^\dagger(\bx)
      \phi^\dagger(\by)\psi(\by)\phi(\bx)
  - \psi^\dagger(\bx)\rho(\by)\psi(\bx)\right].~~~~
\label{ham1}
\end{eqnarray}
Note that for 
\begin{equation}
  \bW(\bx) = \mbox{\boldmath $\nabla$}\times\ln|\bx|
\end{equation}
we get $\mbox{\boldmath $\nabla$}\times{\bf W}(\bx) =-2\pi\delta(\bx)$, 
and the Hamiltonian
reduces to the self-dual form
\begin{equation}
 H_{\rm SD} = {1\over 2m}\int d\bx\,\left[|\D^+\phi(\bx)|^2 + 
       |\D^+\psi(\bx)|^2\right] \; .
\end{equation}
Alternatively, the Hamiltonian can be expressed in terms of the vector
\begin{equation}
 \mbox{\boldmath $\D$} = \mbox{\boldmath $\nabla$} 
   - i\int d\by \bW(\by - \bx)\rho(\by) 
\end{equation}
as
\begin{eqnarray}
 \lefteqn{H = {1\over 2m}\int d\bx\,\left[|\mbox{\boldmath $\D$}\phi|^2 
                   + |\mbox{\boldmath $\D$}\psi|^2\right]
     + {1\over 2m}\int d\bx\,d\by\,
      \mbox{\boldmath $\nabla$}_y\times{\bW}(\by-\bx)}
     \nonumber\\
 & & \qquad\qquad \qquad \mbox{} \times
      \Big[:\big(|\phi(\bx)|^2 - |\psi(\bx)|^2\big)\rho(\by):  
        + 2\psi^\dagger(\bx)\phi^\dagger(\by)\psi(\by)\phi(\bx)\Big].
     \qquad\qquad
\label{ham2}
\end{eqnarray}
Note that this Hamiltonian possesses a local symmetry given by
\begin{eqnarray}
 \bW &\longrightarrow & \bW + \mbox{\boldmath $\nabla$} f \nonumber \\
 \phi,\psi(\bx) &\longrightarrow & \phi,\psi(\bx)\exp{\left[
       i\int d\by f(\by-\bx)\rho(\by)\right]} \; ,
\label{symmetry}
\end{eqnarray}
which suggests identifying {\boldmath $\D$} with a covariant derivative in a background
abelian gauge field given by
\begin{equation}
    {\bf A}(\bx) = \int d\by\,{\bf W}(\by-\bx)\rho(\by) \; .
\end{equation}
The above symmetry is then simply the gauge invariance
\begin{equation}
  {\bf A} \longrightarrow {\bf A} + \mbox{\boldmath $\nabla$}\Lambda\; ,
\end{equation}
with $\Lambda = \int d\by f(\by-\bx)\rho(\by)$,
and the Hamiltonian in (\ref{ham1}) or (\ref{ham2}) is interpreted
as minimal coupling to the gauge field plus non-minimal couplings.
We will see that for a specific choice (up to the local symmetry in 
(\ref{symmetry})) of the superpotential $\bW$ this theory reduces
to a Super-Galilean gauge theory.

\section{N-R Super-Chern-Simons Theory}
\setcounter{equation}{0}
\noindent
Gauge theories in odd spacetime 
dimensions with the so called Chern-Simons kinetic term are
topological, and therefore invariant under all coordinate transformations,
including both the Poincar\'e group and the Galilei group. When coupled to
non-relativistic matter only the Galilei group survives, and one ends
up with a non-relativistic (Galilei invariant) gauge theory. 

The action for $2+1$-dimensional Chern-Simons theory coupled to 
non-relativistic matter
is given by 
\begin{eqnarray}
 S_{\rm CS} &=& 
   \int d^3x\,\bigg[ {\kappa\over 2}\partial_t{\bf A}\times{\bf A} - 
   \kappa A^0 B + \phi^\dagger\Big(i\D_t + {\D^2\over 2m}\Big)\phi 
   + \psi^\dagger\Big(i\D_t + {\D^2\over 2m}\Big)\psi \qquad \nonumber\\
  && \qquad \quad \mbox{} - {e\over 2m}B|\psi|^2 + \lambda_1 |\phi|^4 + \lambda_2 |\phi|^2 
  |\psi|^2\bigg] \; .
\label{csaction}
\end{eqnarray}
The Pauli interaction has been included explicitly, as well as the two possible
non-minimal quartic interactions with coefficients $\lambda_1,\lambda_2$. Higher
power terms have been suppressed. 

In addition to Galilei invariance, this action also possesses 
an $SO(2,1)$ conformal symmetry,\cite{leblanclozano} 
which is however broken quantum
mechanically.\cite{bergmanlozano} For the following values of the coupling
constants 
\begin{equation}
 \lambda_1 = - {1\over 2m\kappa} \qquad, \qquad\lambda_2 = 3\lambda_1 
\label{critical}
\end{equation}
the theory is supersymmetric (under $\stwog$),\cite{leblanclozano} and
the conformal anomaly vanishes.\cite{bergmanlozano} This is N-R Super-Chern-Simons
theory. The Hamiltonian of this theory is easily derived from the action
(\ref{csaction}) at the critical point (\ref{critical}),
\begin{equation} 
 H_{\rm SCS} = {1\over 2m} \int d\bx\,\left[|\D\phi|^2 + |\D\psi|^2
      + {1\over\kappa}:|\phi|^4: 
      + {2\over\kappa} |\phi|^2 |\psi|^2\right] \; .
\end{equation}
The gauge field is completely determined by Gauss' law, and is given in 
Coulomb gauge by
\begin{equation}
 {\bf A}(\bx) = -{1\over\kappa} 
      \int d\by\,\left[\mbox{\boldmath $\nabla$}_y\times\ln|\by-\bx|\right]\rho(\by)\; .
\end{equation}
Consequently we see that our Hamiltonian (\ref{ham2}) reduces to $H_{\rm SCS}$
when ${\bf W}(\bx) = -(1/\kappa)\mbox{\boldmath $\nabla$}\times\ln|\bx|$. 
As a bonus we see
that since this is precisely the superpotential for which $H$ was self-dual, we
have
\begin{equation}
 H_{\rm SCS} = H_{\rm SD} \; ,
\end{equation}
in agreement with Leblanc et. al.\cite{leblanclozano} All this suggests
an interesting connection among $d=2$ Galilean supersymmetry, self-duality
and $SO(2,1)$ conformal symmetry, which may or may not go beyond this
Chern-Simons example.

\section{Matrix Fields}
\renewcommand{\theequation}{\arabic{sectionc}.\arabic{equation}}
\setcounter{equation}{0}
\noindent
A composite formulation of superstring theory was mentioned as 
one of the motivations for studying Super-Galilei invariant field
theories.  The main idea
of superstring-bit models is to replace the free string
by a chain of particles
with dynamics given by {\em nearest-neighbor} interactions,
\begin{equation}
 h = \sum_{k=1}^N\left[{\bp_k^2\over 2m} + V(\bx_{k+1}-\bx_k)\right] \; .
\end{equation}
Second-quantization of this system requires the use of matrix-valued
fields.\cite{thornbits} We are led to a Super-Galilei invariant $N_c\times N_c$
matrix field theory, whose
\begin{itemize}
 \item $N_c\rightarrow\infty$ limit yields the nearest-neighbor interactions,
       which become the free string in the continuum limit.
 \item $1/N_c$ expansion produces the chain splitting and joining
       processes which become the string interactions in the continuum limit.
\end{itemize}
The fields transform in the adjoint representation of a
global $U(N_c)$  "color" group, and satisfy canonical 
(anti-)commutation relations,
\begin{equation}
 [\phi(\bx)_\alpha^\beta,\phi^\dagger(\by)_\gamma^\delta] = 
  \{\psi(\bx)_\alpha^\beta,\psi^\dagger(\by)_\gamma^\delta\} = \delta(\bx-\by)
  \delta_\alpha^\delta\, \delta_\gamma^\beta \; .
\end{equation}

\subsection{The Supercharges}
\noindent
The generators of $\stwog$ are singlets, and are given by traces of
products of matrix fields. The one-body (free) operators can be obtained
from their single-component field-theoretic counterparts (\ref{freefield})
by elevating the fields to matrices and tracing, e.g.
\begin{equation}
  R_0 = {1\over 2\sqrt{m}}
      \int d\bx\, \Tr\left[\psi^\dagger(\bx)\partial^+\phi(\bx)\right] \; .
\end{equation}
For the two body operators ($R^\prime, H$) this procedure is ambiguous, since
different (non-cyclic) matrix orderings yield different traces.  In addition,
one is not guaranteed  that a particular ordering will satisfy the superalgebra.
Consider the two possibilities for ordering the matrix fields in $R^\prime$ 
(\ref{Rprime}):
\begin{eqnarray}
R_1^\prime &=& \int d\bx\,d\by \,W^+(\by-\bx):\Tr\left[
          \psi^\dagger(\bx)\rho(\by)\phi(\bx)\right]: \nonumber\\
R_2^\prime &=& \int d\bx\,d\by \,W^+(\by-\bx):\Tr\left[
          \psi^\dagger(\bx)\phi(\bx)\rho(\by)\right]: \; .
\end{eqnarray}
These have the same (normal) ordering of operators, but different color 
correlations.  The corresponding supercharges are given by $R_1=R_0 + R_1^\prime$
and $R_2 = R_0 +  R_2^\prime$, neither of which satisfy the superalgebra, since:
\begin{equation}
 \{R_1,Q^\dagger\} \neq   -P^+/2  \quad {\mbox{for all }} W^+(\bx)\neq 0
 \qquad , \qquad
 \{R_2,R_2\}   \neq  0 \; .
\end{equation}
However, the second anti-commutator involves the factor
\begin{equation}
 W^+(\bx-\bz)W^+(\bz-\by)+W^+(\bx-\by)W^+(\by-\bz)+
     W^+(\bz-\bx)W^+(\bx-\by)\; ,
\end{equation}
which vanishes for  $W^+(\bx) = \alpha\,\partial^+\ln|\bx| $, 
or in vector notation:
\begin{equation} 
  \bW(\bx) = \alpha_1\mbox{\boldmath $\nabla$}\ln|\bx|
    + \alpha_2\mbox{\boldmath $\nabla$}\times\ln|\bx|  \qquad {\rm where} \quad
  \alpha = \alpha_1 - i\alpha_2 \; .
\label{superpot}
\end{equation}
With this superpotential, $R_2$ satisfies the $\stwog$ algebra, and the 
field theory defined
by $H = 2\{R_2, R_2^\dagger\} $
is $\stwog$-invariant.

\subsection{Large $N_c$ and Closed Chains}
\noindent
The Fock space contains states transforming under various
irreducible representations of $U(N_c)$, but we are mostly
interested in the singlet closed-chain states:
\begin{equation}
 \ket{\Psi}=\int\prod_{k=1}^N\big(d\bx_kd\theta_k\big)
 \Tr[\Phi^\dagger({\bf x}_1,\theta_1)\cdots\Phi^\dagger({\bf x}_N,\theta_N)]\ket{0}
 \Psi({\bf x}_1,\theta_1,\cdots,{\bf x}_N,\theta_N),
\end{equation}
where $\Phi^\dagger(\bx,\theta) = \phi^\dagger(\bx) + \psi^\dagger(\bx)\theta$,
and $\Psi$ is a many-body wavefunction.
The action of two-body operators on this state will give rise to terms with
two traces, corresponding to splitting the chain into two chains. Operators
with color-correlated annihilation operators will also produce
single trace terms, preserving the integrity of the chain. These will
be enhanced by a factor of $N_c$ relative to the chain-splitting terms. To see
how this works in a simple setting consider a matrix field $a(x)$. The
kinds of two-body operators we are dealing with are:
\begin{eqnarray}
 \Omega_1 &=&{1\over N_c}\int dxdyV(y-x)
    \Tr\left[a^\dagger(x)a^\dagger(y)a(y)a(x)\right] \nonumber\\
 \Omega_2 &=& {1\over N_c}\int dxdyV(y-x)
   :\Tr\left[a^\dagger(x)a(y)a^\dagger(y)a(x)\right]: \; .
\end{eqnarray}
In $\Omega_1$ the annihilation operators are correlated, and in $\Omega_2$ 
they are not. Acting on a single $N$-particle chain 
$\ket{\psi} =\Tr\left[a^\dagger(x_1)\cdots a^\dagger(x_N)\right]\ket{0}$,
gives after one contraction \goodbreak
\begin{eqnarray}
 \lefteqn{\Omega_1\ket{\psi}={1\over N_c}\int dy\sum_k V(y-x_k)\times} \nonumber\\
 & & \times\Tr\big[ a^\dagger(x_k)a^\dagger(y)a(y)a^\dagger(x_{k+1})\cdots 
   a^\dagger(x_N)a^\dagger(x_1)\cdots a^\dagger(x_{k-1})\big]\ket{0}~~~~~ \nonumber\\[10pt]
 \lefteqn{\Omega_2\ket{\psi}={1\over N_c}\int dy\sum_k V(y-x_k)\times}\nonumber\\
 & & \times \Tr\big[ a^\dagger(x_k):a(y)a^\dagger(y):a^\dagger(x_{k+1})\cdots 
   a^\dagger(x_N)a^\dagger(x_1)\cdots a^\dagger(x_{k-1})\big]\ket{0} \; .~~~~~ 
\end{eqnarray}
The last contraction of $a(y)$ with one of the creation operators to its right
will almost always break the trace into two traces. The one exception is
when $a(y)$ contracts with $a^\dagger(x_{k+1})$ in the action of $\Omega_1$,
giving the original trace with a factor of $N_c$ coming from the
trace of the identity. 
In the limit $N_c\rightarrow\infty$ we therefore get
\begin{equation}
 \Omega_1\ket{\psi} = \sum_{k=1}^N V(x_{k+1}-x_k)\ket{\psi} \quad , \quad
 \Omega_2\ket{\psi} = 0 \; ,
\end{equation}
exhibiting nearest-neighbor interactions for correlated annihilation
operators.

The supercharge $R_2$ (and thus $H$) will
not give rise to nearest-neighbor interactions, since its annihilation
matrices are uncorrelated. On the other hand $R_1$ will, but 
it failed to satisfy the superalgebra. 
Both properties are required to build supersymmetric chains. It turns out that the 
combination\footnote{The (anti-)commutators apply only to matrix ordering.}
\begin{equation} 
 R^\prime = \int 
       W^+(\by-\bx):\Tr\bigg[\Big(
       \left[\phi^\dagger(\by),\phi(\by)\right] 
       + \left\{\psi^\dagger(\by),\psi(\by)\right\}\Big)
       \left[\psi^\dagger(\bx),\phi(\bx)\right]\bigg]: \; ,
\end{equation}
does precisely this. Again we find that the superalgebra requires
the superpotential (\ref{superpot}).
Note that this superpotential  is the same as the one
that gave N-R Super-Chern-Simons theory in the previous section, up to
a local symmetry transformation (\ref{symmetry}). This seems to imply that 
the Super-Galilean matrix field theory is related to {\em non-abelian}
N-R Super-Chern-Simons theory. This issue is under current 
investigation.\cite{bergmandunne}

\subsection{Chain Quantum Mechanics}
\noindent 
In the limit $N_c\rightarrow\infty$ the supercharge $R=R_0 + R^\prime$ preserves 
the integrity 
of the chain, and gives rise to a first-quantized Hamiltonian given by
\begin{eqnarray}
h &=& {1\over 2m}\sum_{k=1}^N\Bigg\{
        - \nabla_k^2 
       + 2i{\bf W}_{k,k+1}
       \cdot\big(\mbox{\boldmath $\nabla$}_{k}-\mbox{\boldmath $\nabla$}_{k+1}\big)
       + 2i   \partial_k^+W^-_{k,k+1} \nonumber\\
   && \qquad \qquad \mbox{} - i\Big[\partial_k^+W^-_{k,k+1} -
      \partial_k^-W^+_{k,k+1}\Big]
       \big(\theta_{k+1}-\theta_k\big)\left({\partial\over\partial\theta_{k+1}}
      - {\partial\over\partial\theta_k}\right)\nonumber\\
  && \qquad \qquad \mbox{} + 2|{\bf W}_{k,k+1}|^2 
      - W^+_{k-1,k}W^-_{k,k+1}
          - W^+_{k,k+1}W^-_{k-1,k} 
      \Bigg\} \; ,
\end{eqnarray}
where the shorthand notation $W_{k,k+1}\equiv W(\bx_k-\bx_{k+1})$ was used.
We wish to determine whether chains really form as bound states in this theory. 
Since $\alpha$ is a dimensionless (complex) parameter, this system
is classically scale invariant, and it seems unlikely
that any finite energy bound state should form unless an anomaly appears.
We will show that there is no
anomaly, and scale invariance is exact. 
This has the unfortunate effect of ruling out this theory as a 
superstring-bit model.

Take a closer look at the two-body sector, namely what would be a single
link in the chain. The two-body wavefunction has four components:
\begin{equation}
u_1(\bx_1,\bx_2)
  + (\theta_1+\theta_2)u_2(\bx_1,\bx_2)
  + (\theta_1-\theta_2)u_3(\bx_1,\bx_2)
  + \theta_1\theta_2u_4(\bx_1,\bx_2) \; , 
\end{equation}
and the relative coordinate space Hamiltonian is given by 
\begin{equation}
  h_2 = -{\nabla^2\over m} 
   + {1\over mx^2}\left[2i\alpha_1\bx\cdot\mbox{\boldmath $\nabla$}
         - 2i\alpha_2\bx\times\mbox{\boldmath $\nabla$} + |\alpha|^2\right] 
  + {2\pi i\over m}(\alpha_1\pm i\alpha_2)\delta^{(2)}(\bx) \; .
\label{twobody}
\end{equation}
The upper and lower signs in the coefficient of the $\delta$-function
hold for $u_1,u_2$ and $u_3,u_4$, respectively.
$\alpha_1$ can be eliminated by a phase redefinition 
\begin{equation}
 u_n(\bx)  = x^{-i\alpha_1} \tilde{u}_n(\bx) \; ,
\end{equation}
corresponding to a "gauge" transformation (\ref{symmetry}), 
with $f(\bx) = \alpha_1\ln|\bx|$. This leaves precisely 
the two-body Hamiltonian arising from
the self-dual N-R Matter-Chern-Simons theory,\cite{dunnebook}
which is known to be scale invariant to all orders in 
perturbation theory.\cite{bergmanlozano,freedman,cklee}
There can thus be no finite energy bound state. It has in fact been
shown that the bound state energy vanishes in this case.\cite{bergmantgal}
A very simple argument can be made utilizing supersymmetry. Since
$h = \{r,r^\dagger\}$,
the Hamiltonian is a positive definite operator, which forbids a negative
energy bound state. A positive energy bound state is likewise forbidden
since the potential vanishes at infinity. Therefore the bound state energy
must vanish.\footnote{
The above argument is known to fail
for certain singular superpotentials,\cite{jevicki,nam}
which develop negative energy bound states. 
It seems unlikely however that this happens here, since the Hamiltonian
(\ref{twobody}) with $\alpha_1=0$ is equivalent to the Aharonov-Bohm
problem, for which the exact solution is scale invariant.}

\section{Discussion}
\renewcommand{\theequation}{\arabic{sectionc}.\arabic{equation}}
\setcounter{equation}{0}
\noindent
Since the Galilei group (and its extension $\stwog$) can be understood
both as a N-R limit and as a light-cone subgroup of Poincar\'e groups,
it is relevant in the description of both non-relativistic 
(condensed matter) systems,
and relativistic systems in an infinite momentum (light-cone) frame.
We have stressed the latter point of view, as it applies to 
reformulating string theory in terms of constituents.

Quartic $2+1$-dimensional $\stwog$-invariant field theories are defined by
a vector superpotential given by
\begin{equation}
 \bW(\bx) = \mbox{\boldmath $\nabla$}f(|\bx|) + 
           \mbox{\boldmath $\nabla$}\times g(|\bx|) \; ,
\end{equation}
where $f(|\bx|)$ makes no contribution to the dynamics.
N-R Super-Chern-Simons theory emerges as a special case when 
$g(|\bx|) = \ln|\bx|$. If the fields are matrices,
$\stwog$-invariance requires
\begin{equation}
 \bW(\bx) = \alpha_1\nabla\ln|\bx| + \alpha_2\nabla\times\ln|\bx|\; ,
\end{equation}
suggesting an interpretation in terms of non-abelian 
N-R Super-Chern-Simons theory. This in turn implies an exact conformal
symmetry, precluding a finite energy bound state. If the particles
do not bind into chains, the continuum limit cannot be a string theory.
It is still possible that the chains are zero-energy bound states,
but then it isn't clear how the string tension will arise.
To achieve finite energy bound chains in an 
$\stwog$-invariant matrix theory,
it may be necessary to beyond quartic terms in the supercharge.

\nonumsection{Acknowledgments}
\noindent
I would like to thank the organizers of the workshop for
a stimulating meeting. 
This work was supported in part
by the Department of Energy under grant DE-FG05-86ER-40272,
by the Institute for Fundamental Theory, and
by the National Science Foundation under grant PHY-9315811.

\nonumsection{References}

\end{document}